\documentclass[aps,prl,twocolumn,superscriptaddress]{revtex4-2}
\usepackage{amsmath,amssymb}
\usepackage{graphicx}
\usepackage[colorlinks=true,citecolor=blue,linkcolor=blue,urlcolor=blue]{hyperref}

\usepackage{amsmath,graphicx,array}
\usepackage{dcolumn,soul}%

\usepackage{amsthm}
\usepackage{algorithm, algorithmicx, algpseudocode}
\usepackage{listings}%
\usepackage{hyperref}
\usepackage{braket}
\usepackage{upgreek}
\usepackage{booktabs}
\usepackage{array} 
\hypersetup{
colorlinks=true,
linkcolor=blue,
filecolor=blue,
urlcolor=magenta,
citecolor=blue,
}
  
\def\uns{\ifmmode\,\else$\,$\fi}%

\begin{document}
	
	% \dhead{RESEARCH ARTICLE}
	
	% \subhead{PHYSICS}
	
	\title{High-fidelity two-qubit 
quantum logic gates in a trapped-ion chain using axial motional modes}
	
	\author{Xingyu Zhao}
\thanks{These authors contributed equally to this work.}
\affiliation{Laboratory of Spin Magnetic Resonance, School of Physical Sciences, Anhui Province Key Laboratory of Scientific Instrument Development and Application, University of Science and Technology of China, Hefei, 230026, China}
\affiliation{Hefei National Laboratory, University of Science and Technology of China, Hefei, 230088, China}

\author{Ji Bian}
\thanks{These authors contributed equally to this work.}
\affiliation{Laboratory of Spin Magnetic Resonance, School of Physical Sciences, Anhui Province Key Laboratory of Scientific Instrument Development and Application, University of Science and Technology of China, Hefei, 230026, China}

\author{Yi Li}
\affiliation{Laboratory of Spin Magnetic Resonance, School of Physical Sciences, Anhui Province Key Laboratory of Scientific Instrument Development and Application, University of Science and Technology of China, Hefei, 230026, China}
\affiliation{Hefei National Laboratory, University of Science and Technology of China, Hefei, 230088, China}
\affiliation{National Advanced Talent Cultivation Center for Physics, University of Science and Technology of China, Hefei, 230026, China}

\author{Yue Li}
\affiliation{Laboratory of Spin Magnetic Resonance, School of Physical Sciences, Anhui Province Key Laboratory of Scientific Instrument Development and Application, University of Science and Technology of China, Hefei, 230026, China}

\author{Mengxiang Zhang}
\affiliation{Anhui Province Engineering Research Center for Quantum Precision Measurement, University of Science and Technology of China, Hefei, 230088, China}

\author{Yiheng Lin}
\email{yiheng@ustc.edu.cn}
\affiliation{Laboratory of Spin Magnetic Resonance, School of Physical Sciences, Anhui Province Key Laboratory of Scientific Instrument Development and Application, University of Science and Technology of China, Hefei, 230026, China}
\affiliation{Hefei National Laboratory, University of Science and Technology of China, Hefei, 230088, China}

\date{\today}

\begin{abstract}
Trapped-ion systems are one of the leading platforms for quantum information processing, where a key challenge is to scale up system size while maintaining high-fidelity two-qubit operations. A promising approach is to build high-performance modules interconnected via strong coupling. In particular, axial motional modes provide a feasible means of coupling short ion chains. However, previous implementations of fully connected 5-ion modules based on axial modes have been limited to fidelities of $96.6–98.0\%$. Here, we demonstrate two-qubit quantum logic gates in a 5-ion $^{40}$Ca$^{+}$ chain using axial modes, achieving fidelities exceeding 99\% for adjacent pairs and over 98\% for arbitrary pairs by carefully tackling dominant error sources. Our results are beneficial to the development of scalable ion-trap quantum processors, quantum simulation and quantum-enhanced metrology.
\end{abstract}
	%\jelcode{Pa, J6, P16, E22}
	
	%\keywords{Quantum information processing, Trapped ion, Quantum logic gates}

	\maketitle
\section{Introduction}
\label{sec:1}

Trapped-ion systems are among the most promising platforms for quantum computing, owing to their exceptional gate fidelities, long coherence times, and potential for scalability~\cite{Bru, Pen, WuY}. For single-ion or ion-pair systems, both single- and two-qubit gate fidelities have surpassed the $99.9\%$ fault-tolerance threshold~\cite{Gae, Bal, Cla}. Considerable progress has been made toward direct scaling via long linear ion chains\cite{Wri, Pog, Cet, Pos1, Pos2}, including demonstrations of all-to-all two-qubit operations based on radial modes with average fidelities of $99.5\%$ across 30 ions~\cite{Jwo}. Alternatively, two-dimensional (2D) ion crystals offer another path to scaling~\cite{Guo}, enabling entangling gates between arbitrary pairs in a 4-ion array with fidelities ranging from $96.0\%$ to $98.6\%$~\cite{Hou}.  However, as system size increases, maintaining high-fidelity two-qubit gates becomes increasingly challenging due to control complexity and noise accumulation. This imposes a fundamental bottleneck on direct scaling approaches. A natural solution is to adopt a modular architecture, dividing complexity into intra-module and inter-module operations, which can be optimized independently~\cite{Mon}. Notable modular strategies for trapped ions include quantum charge-coupled devices (QCCD) based on ion shuttling~\cite{Kie, Pin, Mos}, and distributed processors using ion-photon networks~\cite{Moeh, Dua, Ste, Fen, Mai}. QCCD architectures scale by transporting ions and have demonstrated high-fidelity ($\textgreater99.8\%$) two-qubit gates in a 56-qubit processor~\cite{decross}. Its practical implementations need to overcome several challenges, such as complex operations including ion separation, shuttling, recombination, and re-cooling, resulting in a duty cycle below $2\%$ for quantum gate operations~\cite{Mos}. %Distributed processors using ion-photon interfaces have achieved deterministic teleportation of a controlled-Z gate between separate modules with $86\%$ fidelity, but are limited by the slow entanglement generation rate, requiring $\sim8.3$ ms per successful attempt~\cite{Mai}. 

 \begin{figure*}[ht]
\centering
\makebox[\textwidth][c]{\includegraphics[width=13.5cm]{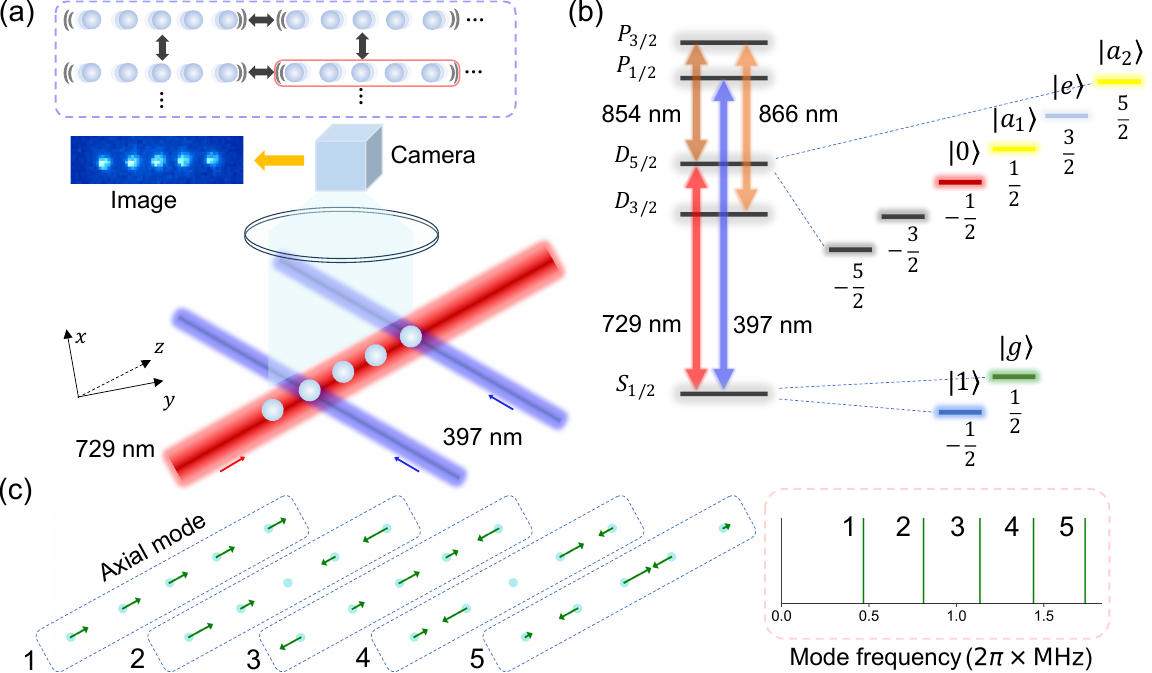}}\\
\caption{
(a) Schematic of the quantum spring array (QSA) architecture (top) and our experimental setup (bottom). 
In the QSA architecture, modular short ion chains—each comprising a small number (e.g., five) of fully controllable qubits—serve as independent modules. These modules can be transported collectively, and inter-module entangling operations are enabled via axial-mode coupling. Axial mode coupling occurs along both the axial and radial directions, allowing for scalable two-dimensional trapped-ion quantum processors. In our experimental setup, ions are manipulated by a global 729~nm laser aligned along the axial direction and individually addressed by a 397~nm laser along the radial direction. Spin-state readout is performed via site-resolved imaging using a camera.  
(b) Energy-level diagram of the $^{40}$Ca$^{+}$ ion. The qubit is encoded in the states \( \ket{0} \) and \( \ket{1} \), while \( \ket{a_{1}} \), \( \ket{a_{2}} \), \( \ket{g} \), and \( \ket{e} \) are auxiliary states used for optical pumping and state shelving.  
(c) Mode vectors and frequency spectrum of the axial motional modes. The length of each vector represents the mode participation (coupling strength) of the corresponding ion for that mode.
}

\label{fig1}
\end{figure*}

An alternative inter-module connection involves direct Coulomb coupling between ions in separate potential wells~\cite{Har, Bro}. Each well contain multiple ions and form a module. This method has enabled entanglement between two ions in adjacent wells with $82\%$ fidelity~\cite{Wil}. Crucially, axial motional modes provide stronger inter-module coupling than radial modes under typical trapping conditions~\cite{Wil, james}, with the coupling strength scaling nearly quadratically with ion number~\cite{Val}. For example, in 5-ion chains, axial-mode coupling can be an order of magnitude stronger than radial coupling. 
Moreover, coupling of axial modes along both axial and radial directions have been demonstrated, supporting the feasibility of a two-dimensional modular architecture recently proposed, the \textit{quantum spring array} (QSA)~\cite{Val}, as illustrated in Fig.~\ref{fig1}(a). The QSA employs short ion chains as fully controllable modules, which are collectively shuttled and coupled via axial modes to enable inter-module entangling operations. Thus compared to QCCD, QSA potentially reduces technical complexity by minimizing shuttling overhead. 
Moreover, axial modes in short chains offer advantages such as a sparse mode spectrum and strong coupling. However, challenges including individual control and noise suppression have limited axial-mode-based gate fidelities. Previous demonstrations of fully connected 5-ion axial-mode-based gates achieved fidelities of $96.6\%–98.0\%$~\cite{Manovitz}. Improving intra-module axial-mode entangling gates is thus crucial.

 In this work, we demonstrate high-fidelity entangling gates between arbitrary ion pairs in a 5-ion $^{40}$Ca$^{+}$ module using axial modes. We achieve fidelities exceeding $99\%$ for adjacent pairs and over $98\%$ for non-adjacent pairs by systematically identifying and suppressing major error sources. 
These results can be directly integrated with QSA to realize scalable, efficient building blocks for large-scale ion-trap quantum processors. They also benefit applications in quantum simulation~\cite{Lan, Kus} and quantum-enhanced metrology~\cite{ddp, ito}.

\section{Experimental Realization}
\label{sec:2}

The experiment is performed using a segmented blade trap that confines a linear chain of five $^{40} \mathrm{Ca}^{+}$ ions along the axial direction, as depicted in Fig.~\ref{fig1}(a). Each ion encodes an optical qubit using the electronic states \( \ket{1} = \ket{^{2}S_{1/2}, m_j = -1/2} \) and \( \ket{0} = \ket{^{2}D_{5/2}, m_j = -1/2} \) [see Fig.~\ref{fig1}(b)]. The axial motional mode spectrum spans 0.46–2.07~MHz, with inter-ion spacings ranging from 5.40 to 6.75~$\upmu$m. A typical axial mode configuration used in the experiment is shown in Fig.~\ref{fig1}(c) (see Supplementary Material for details).

A narrow-linewidth 729~nm laser beam aligned along the trap axis uniformly illuminates all five ions, coherently driving quadrupole transitions between the \( S_{1/2} \) and \( D_{5/2} \) manifolds. This beam path supports both resonant carrier transitions for single-qubit rotations and spin-motion coupling for two-qubit entangling gates.

To implement entangling gates between arbitrary ion pairs, we adopt the Mølmer–Sørensen (MS) protocol~\cite{MSo1,MSo2,MSo3,MSo4}, which generates effective spin–spin interactions via shared motional modes. A global 729~nm beam with bichromatic modulation—consisting of two tones symmetrically detuned from the qubit resonance by approximately the motional frequency—produces a spin-dependent force. By selecting a specific axial mode and tuning the detuning appropriately, the spin-dependent displacements trace closed trajectories in phase space at gate times, yielding entangling gates on the targeted ion pair (see Supplementary Material).

\begin{figure}[ht]
\centering
\makebox[\linewidth][c]{\includegraphics[width=6.75cm]{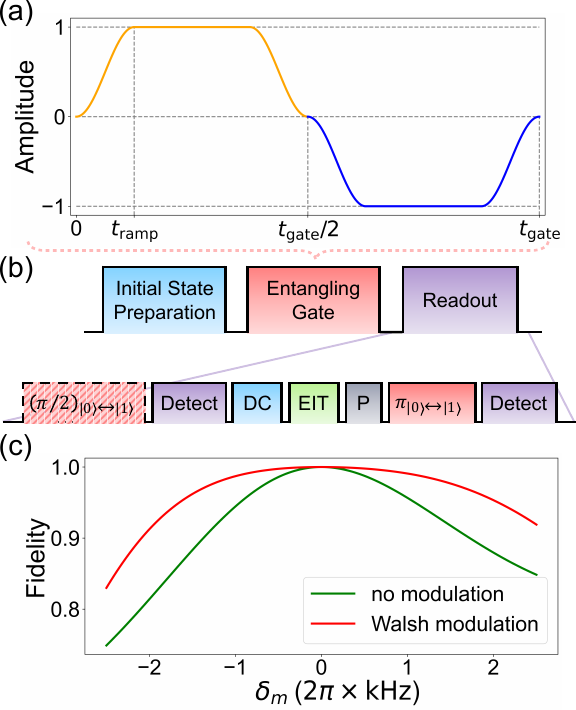}}\\
\caption{(a) Control pulse shape for the entangling gate, featuring Walsh modulation (sign inversion) and ramped amplitude profiles. The negative amplitude corresponds to a $\pi$ phase shift of the 729~nm bichromatic laser field. (b) Experimental sequence including initial state preparation, entangling gate operation, and readout. The readout pulse sequence consists of a $\pi/2$ analysis pulse (only for parity measurement), first fluorescence detection, Doppler cooling (DC), electromagnetically induced transparency (EIT) cooling, optical pumping (P),  $\pi$-pulse, and second fluorescence detection. (c) Simulated fidelity of the MS gate with and without Walsh modulation under detuning mismatch $\delta_m$, showing improved robustness to trap frequency noise.}
\label{fig2}
\end{figure}

In addition to global control, site-resolved operations are required to select the target ion pair. This is achieved using tightly focused 397~nm beams propagating radially, which are steered via two perpendicular acousto-optic deflectors (AODs)~\cite{Pog,cAOD}, allowing dynamic individual addressing with uniform frequency.

The experimental sequence comprises three main stages [Fig.~\ref{fig2}(b)]: (1) initial state preparation, (2) entangling gate, and (3) readout. Motional initialization is accomplished via Doppler cooling (DC), electromagnetically induced transparency (EIT) cooling, and sideband cooling (SBC), bringing all axial modes near the ground state to suppress the Debye–Waller effect~\cite{Bib}. All ions are optically pumped to \( \ket{g} = \ket{^{2}S_{1/2}, m_j = 1/2} \) through a two-step cycle: a 729~nm $\pi$-pulse transferring \( \ket{1} \to \ket{e} = \ket{^{2}D_{5/2}, m_j = 3/2} \), followed by repumping using 854 and 866~nm lasers.
Next, a selected ion pair is transferred to \( \ket{1} \) using co-propagating Raman beams~\cite{Gae}. The typical transfer error per ion is on the order of  \( 10^{-2} \), with negligible crosstalk (\(< 2 \times 10^{-6}\)) on neighboring ions (see Supplementary Material). Spectator ions are shelved into auxiliary sublevels \( \ket{a_1}, \ket{a_2} \in D_{5/2} \) via two global 729~nm $\pi$-pulses, effectively isolating the computational subspace.

To achieve high-fidelity entangling gates, we identify and mitigate two major error sources. First, Kerr nonlinearity induces coupling between thermal radial phonons and axial mode frequencies, resulting in dephasing~\cite{phne}. To suppress this, we implement a first-order Walsh modulation~\cite{Walsh} by inserting a $\pi$ phase flip midway through the gate [Fig.~\ref{fig2}(a)]. This two-loop structure averages out low-frequency detuning noise without additional laser pulses. Figure~\ref{fig2}(c) shows a simulated comparison of gate fidelity versus differential detuning $\delta_m$ deviated from optimal value of the bichromatic fields for gate operation, illustrating improved robustness to motional frequency noise under Walsh modulation.
Second, square pulses for the bichromatic field produce spectral leakage, causing off-resonant excitations. To mitigate this, we shape the pulse with a $\sin^2$ amplitude envelope using rise/fall times $t_{\text{ramp}}$ [Fig.~\ref{fig2}(a)], which narrows the frequency spectrum~\cite{Win}.

Gate performance is characterized by the Bell-state fidelity \(\mathcal{F}_{\text{Bell}}\), obtained via partial state tomography (PST)~\cite{Leib,Hug}:
\[
\mathcal{F}_{\text{Bell}} = \langle \psi_{\text{Bell}}|\rho_{\textrm{f}}|\psi_{\text{Bell}}\rangle = \tfrac{1}{2}(P_{00}+P_{11}) + |\rho_{00,11}|,
\]
where \( \psi_{\text{Bell}} = (\ket{00} + e^{i\varphi} \ket{11})/\sqrt{2} \) , $P_{ij}=\langle ij|\rho_{\textrm{f}}|ij\rangle$ are the two-qubit populations, and \( \rho_{00,11} = \langle 00|\rho_{\textrm{f}}|11\rangle\) is the coherence between \( \ket{00} \) and \( \ket{11} \).
Compared to MS gates in 2-ion or long chains using radial modes, 
our protocol introduces an extra step of selective state transfer, which contributes dominant error due to 397~nm Raman power drifts. To better reveal the intrinsic gate errors, we adopt a post-selection strategy based on a dual-detection single-shot readout~\cite{jeff,Qud1,Qud2}, to alleviate effects from imperfect state preparation.
Each shot includes two site-resolved detections using a scientific CMOS camera, separated by a 729~nm $\pi$-pulse. Such a post-selection configuration ideally selects events that the qubit states are populated, distinguishing from the erroneous events that the target ions remain in the $\ket{g}$ state due to imperfect initialization. Coherence detection is performed by applying a $\pi/2$ analysis pulse with varying phase before the first detection. Each readout collects 397~nm fluorescence for 2.5~ms and classifies ion states as ``bright'' or ``dark'' using threshold discrimination.

To counteract heating from the first detection—which would degrade the second $\pi$-pulse fidelity—we insert a cooling sequence (DC + EIT) and optical pumping (397~nm \( \sigma_{-} \)) between the two detections to reset residual \( S_{1/2} \) population without affecting \( D_{5/2} \) states. An experimental shot is accepted only if the target ions are observed once as ``bright'' and spectator ions are consistently ``dark'', ensuring the target ions reside in the computational subspace. This post-selection scheme effectively removes initialization errors, primarily from 397~nm Raman transfer. 
%and minor leakage or off-resonant excitation of adjacent carrier transitions. 
A detailed validation of this method and the residual readout error analysis is presented in the Supplementary Material.
After post-selection, we extract the two-qubit populations \( P_{00}, P_{01}, P_{10}, P_{11} \) for the selected ion pair \((j,k)\), satisfying the normalization condition \( P_{00}+P_{01}+P_{10}+P_{11}=1 \).
\begin{figure}[ht]
\centering
\makebox[\linewidth][c]{\includegraphics[width=6.75cm]{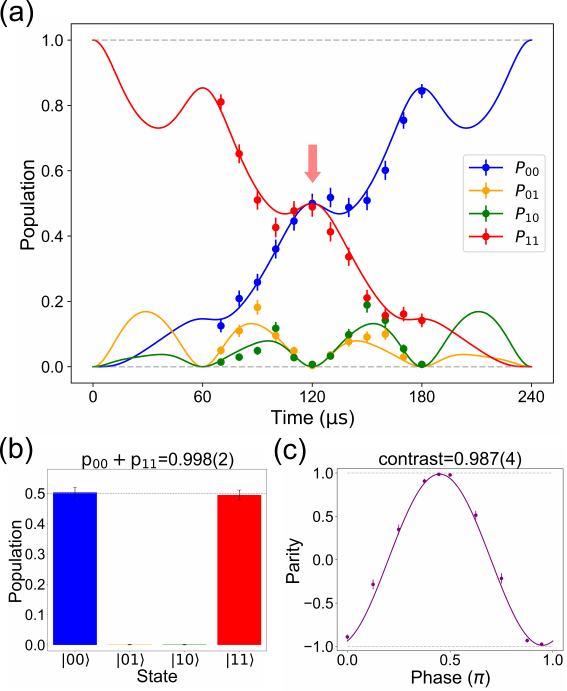}}
\caption{
(a) Time evolution of the populations during the MS gate for ion pair (1,\,2). Each data point represents approximately 300 post-selected repetitions.  
(b) Measured populations of the generated entangled state for ion pair (1,\,2), based on 933 accepted repetitions post-selected from 1000 experimental shots.  
(c) Measured parity oscillation for ion pair (1,\,2), with the contrast extracted by fitting the experimental data after post-selection.  
Error bars in all panels indicate one standard deviation.
}

\label{fig3}
\end{figure}

\begin{figure*}[ht]
\centering 
\makebox[\textwidth][c]{\includegraphics[width=13.5cm]{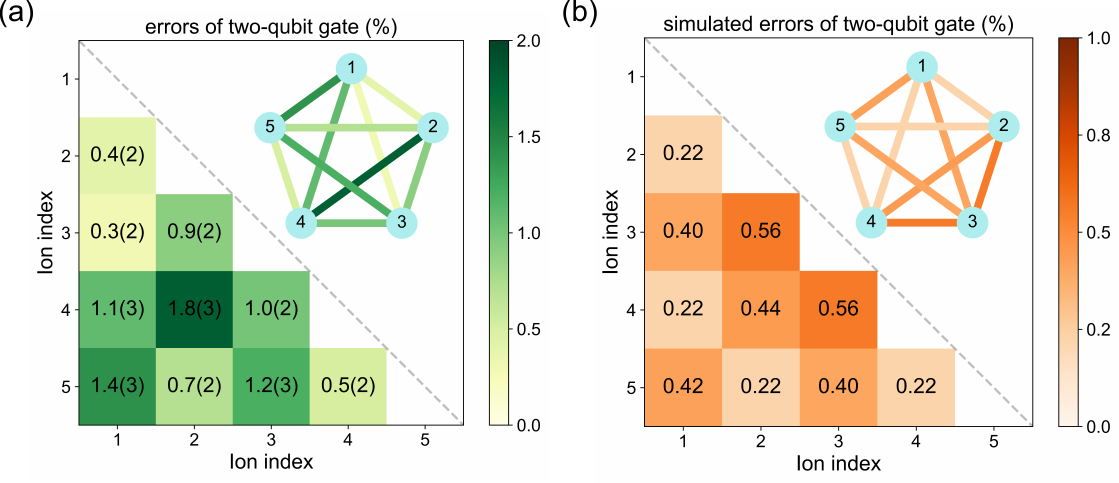}}
\caption{
Two-qubit gate errors for all ion pairs in a fully connected 5-ion chain. 
(a) Experimentally measured gate errors. Fidelities exceed 99\% for adjacent ion pairs and remain above 98\% for all other combinations, demonstrating high-fidelity all-to-all connectivity. 
(b) Numerically simulated gate errors based on independently characterized error sources, mainly including motional mode heating and dephasing, spin dephasing, and Rabi rate fluctuations. The simulations validate the observed gate performance and help identify the dominant error contributions.
}

\label{fig4}
\end{figure*}

\section{Results}

We demonstrate high-fidelity entangling operations between arbitrary ion pairs in a linear 5-ion chain using axial-mode coupling. To illustrate the gate dynamics, we apply the waveform in Fig.~\ref{fig2}(a) to ion pair (1,\,2), abruptly stopping the sequence at time $t$ near $t_{\textrm{gate}}$. The measured populations, shown in Fig.~\ref{fig3}(a), agree with numerical simulations. A high-fidelity Bell state is generated by halting the gate at exactly $t_{\textrm{gate}}$. In our implementation, the gate time is fixed at 120~$\mu$s for most ion pairs, except for pairs (2,\,3) and (3,\,4), which require 140~$\mu$s due to limitations in available laser power. The amplitude ramp duration is set to $t_{\text{ramp}} = 5~\mu\text{s}$ for all gates.

To estimate Bell-state fidelity after a full two-loop gate operation, we first perform 1000 experimental repetitions to measure the population \( P_{00} + P_{11} \) of the Bell state. We then conduct an additional set of experiments with a $\pi/2$ analysis pulse of varying phase to extract the parity \( \Pi = P_{00} - P_{01} - P_{10} + P_{11} \). Due to the AC Stark shift, the parity oscillation exhibits a small phase offset. The parity contrast \( C \) is extracted by fitting the measured data to \( \Pi(\phi) = C\sin{[2(\phi + \phi_0)]}\), where \( \phi_0 \) accounts for the phase bias. We scan 8 evenly spaced phases in the interval $[0, \pi)$, with 300 experimental repetitions per point. To improve fitting accuracy, we include two additional data points near the expected maximum and minimum of the parity oscillation, with 1000 experimental repetitions each. %This refines the fitting precision from roughly 1\% to the order of 0.1\%. 

Figure~\ref{fig3}(b) shows the post-selected populations for pair (1,\,2), with \( P_{00} + P_{11} = 0.998(2) \), and Fig.~\ref{fig3}(c) displays the parity oscillation, yielding a contrast of \( C = 0.987(4) \). Detailed information on the number of accepted repetitions after post-selection is provided in the Supplementary Material. The Bell-state fidelity is calculated as
\[
\mathcal{F}_{\text{Bell}} = \frac{1}{2} \left( P_{00} + P_{11} + C \right) = 0.993(2),
\]
which corresponds to a gate error of \( \epsilon = 1 - \mathcal{F}_{\text{Bell}} = 0.7(2)\% \). Because of post selection, this estimate eliminates preparation errors, but still includes readout imperfections, which contribute an overestimation of 0.3\% (see Supplementary Material). Taking these into account results in an error of $0.4(2)\%$ for pair (1,\,2).
%reconstructed intrinsic gate error for pair (1,\,2) is 

We repeat this benchmarking procedure for all ten ion pairs. As shown in Fig.~\ref{fig4}(a), the gate error remains below 2.0\% for all pairs, and below 1.0\% for all adjacent pairs. These results demonstrate high-fidelity, fully connected two-qubit gate operations. A complete list of fidelities is provided in the Supplementary Material.

To identify performance limitations, we model the dominant error sources using experimentally calibrated parameters. Table~\ref{tab:error_budget} presents the error budget for pair (1,\,2) as an example. The leading contributors are motional mode heating and dephasing, Rabi rate fluctuations, and spin dephasing (detailed discussion available in the Supplementary Material). Figure~\ref{fig4}(b) shows numerical simulations of gate errors for all ion pairs, which agree well with experimental results. Minor deviations are attributed to calibration errors and parameter drifts, since the simulations assume ideal conditions with optimally tuned parameters.

\begin{table}[hb]
    \centering
    \caption{Error budget for ion pair (1,\,2).} 
    \begin{tabular}{p{4cm}p{2cm}<{\centering}}
        \toprule
        Error Source & Contribution ($\times 10^{-3}$)\\
        \midrule
        Rabi rate fluctuations & 0.2 \\
        Spin dephasing & 0.2 \\
        Motional mode dephasing & 0.6\\
        Motional mode heating & 1.1 \\
        Metastable $D_{5/2}$ decay & 0.05\\
        Off-resonant excitation & 0.03\\
        \bottomrule\\
    \end{tabular}
    \label{tab:error_budget}
\end{table}

\section{Conclusion}
\label{sec:4}
In summary, we demonstrate high-fidelity entangling gates based on axial motional modes in a 5-ion $^{40}$Ca$^{+}$ chain, achieving fidelities exceeding 99\% for adjacent qubit pairs and over 98\% for arbitrary pairs. These results are made possible by carefully identifying and suppressing dominant error sources, including Kerr nonlinearity and off-resonant excitations. The laser configuration and spin-phonon coupling scheme employed here are directly transferable to the QSA architecture~\cite{Wil,Val}.
%Our demonstration enables the implementation of arbitrary two-qubit entangling gates within each module composed of a five-ion chain confined along the axial direction. 
The use of uniformly illuminated 729~nm laser beams in addition allows for parallel gate operations across %identical or symmetric ion pairs, 
modules. 
%provided that the trapping parameters are appropriately matched across modules. 
Inter-module entangling gates can be realized by bringing adjacent chains into proximity without merging, leveraging shared axial collective modes to mediate effective spin-spin interactions.
This work establishes a high-fidelity and scalable building block for large-scale trapped-ion quantum processors and offers a promising path forward for applications in quantum simulation and quantum-enhanced metrology~\cite{ddp,ito,Lan,Kus}. Upon submission of our work, we notice recent works utilizing structured light \cite{luyao} and polarization gradient \cite{cui} to induce axial-coupled quantum logic gates in chains of up to six ions, with fidelities of 98.5\% for two ions and up to 97.4\% for more ions. 

\section{Acknowledgment}

\noindent This work was funded by the Innovation Program for Quantum Science and Technology (Grant No.~2021ZD0301603) and the National Natural Science Foundation of China (Grant No.~92165206).

%\section{FUNDING}
%This work was supported by Ministry of Science and Technology of the People’s Republic of China through the National Key R\&D Program of China (2018YFA0306600), Department of Science and Technology of Anhui Province through Anhui Initiative in Quantum Information Technologies (AHY050000) and Anhui Provincial Natural Science Foundation (2108085MA15). 

%\section{AUTHOR CONTRIBUTIONS }
%Yiheng L. conceived the project and guided the experimental investigation and analysis. Xingyu Z. carried out the experiments and performed the numerical analysis. Xingyu Z., Yi L. Yue L. and Mengxiang Z. contributed to building the experimental setup. Xingyu Z. and Bian J. wrote the manuscript.

%\textit{\textbf{Conflict of interest statement. }}None declared.

 % \bibliographystyle{elsarticle-num} 
 % \bibliography{cas-refs}

\onecolumngrid   % 临时切换到单栏
\newpage

\begin{center}
	\bfseries Supplemental Material for\\
	``High-fidelity two-qubit quantum logic gates in a trapped-ion chain using axial motional modes"
\end{center}
\vspace{1em}

%\onecolumngrid   % 切换回双栏

\appendix
% 在补充材料部分开头添加：
\setcounter{figure}{0}
\renewcommand{\thefigure}{S\arabic{figure}}  % 图编号改为 S1, S2...
\setcounter{table}{0}
\renewcommand{\thetable}{S\arabic{table}}    % 表格编号改为 S1, S2...
\setcounter{equation}{0}
\renewcommand{\theequation}{S\arabic{equation}} % 公式编号改为 S1, S2...

\renewcommand{\theHfigure}{S.\arabic{figure}} % 关键点：加 H
\renewcommand{\theHtable}{S.\arabic{table}}
\renewcommand{\theHequation}{S.\arabic{equation}}

% \begin{document}

	\maketitle

\section{I. Mølmer-Sørensen Protocol}

The two-qubit gate between an arbitrary ion pair is implemented using the Mølmer–Sørensen (MS) protocol, which employs a spin-dependent force (SDF) to couple the qubits to collective motional modes. In our experiment, we utilize the axial modes [see Fig.~1(c) in the main text]. The SDF   Hamiltonian for an ion pair \( (j, k) \) coupled to axial mode \( m \) is given by:
\begin{equation}
	\hat{H}_{\mathrm{SDF},m} = \sum_{l \in \{j, k\}} \eta_{l, m} \frac{\hbar \Omega}{2} \hat{a}_m^\dagger e^{-i \delta t} e^{i \phi_l} \hat{\sigma}_x^l + \text{H.c.},
\end{equation}
where $\hat{\sigma}_x^l$ is the Pauli operator on the $l^{\rm  th}$ qubit,  $\hat{a}_m$ ($\hat{a}_m^{\dagger}$) is the annihilation (creation) operator of the $m^{\rm th}$ axial mode, $\delta$ is the frequency detuning between the bichromatic laser field and the motional sideband, $\phi_l$ is the phase associated with mode $m$ of the $l^{\rm th}$ ion,  $\Omega$ is the Rabi rate of the dichromatic laser field, and $\eta_{l,m}$ is the Lamb-Dicke parameter that characterizes the coupling strength between the qubit $l$ and the mode $m$. The SDF with a fixed detuning $\delta$ drives the motional mode along a circular trajectory in phase space in a spin-dependent direction. The unitary evolution under this Hamiltonian (setting \( \hbar = 1 \)) is:
\begin{equation}
	\hat{U}(t) = \exp\left[\sum_{l \in \{j, k\}} \left( \alpha_l(t) \hat{a}_m^\dagger - \alpha_l^*(t) \hat{a}_m \right) \hat{\sigma}_x^l - i \frac{\theta(t)}{2} \hat{\sigma}_x^j \hat{\sigma}_x^k \right],
\end{equation}
where the displacement of qubit $l$ is $\alpha_l(t)=\frac{\Omega \eta_{l,m}\mathrm{e}^{i \phi_l}}{2 \delta}\left(\mathrm{e}^{-i \delta t}-1\right)$, and $\theta(t)=\eta_{j,m}\eta_{k,m}\Omega^2\left(\frac{t}{\delta}-\frac{\sin{(\delta t)}}{\delta^2}\right)$. This operator becomes $\hat{U}_{MS} = \exp{\left(-i \frac{\theta}{2} \hat{\sigma}_x^j \hat{\sigma}_x^k \right)}$ when the condition $\delta t = \pm 2 \pi K, ~K=1, 2, 3,\cdots$ is satisfied. The parameter \( K \) in this condition represents the number of loops. At the end of each loop, the qubit is decoupled from the motional mode. In this experiment, we choose \( K = 2 \) to enable first-order Walsh modulation. The geometric phase $\theta = 2 \pi  \eta_{j,m}\eta_{k,m} K \left(\frac{\Omega}{\delta}\right)^2\mathrm{sign}(\delta)$ corresponds to the area enclosed by the motional trajectory in phase space and represents the phase accumulated between qubits \( (j,k) \) over time. When $\theta = \pm \pi / 2$,  starting from the \( \ket{00} \) state, a maximally entangled Bell state $\left(\ket{00} + e^{i\varphi} \ket{11} \right)/\sqrt{2}$ is generated, where $\varphi = \mp \pi/2$.

\section{II. Selection of Axial Modes}

\begin{figure*}[ht]
	\centering
	\makebox[\textwidth][c]{\includegraphics[width=14cm]{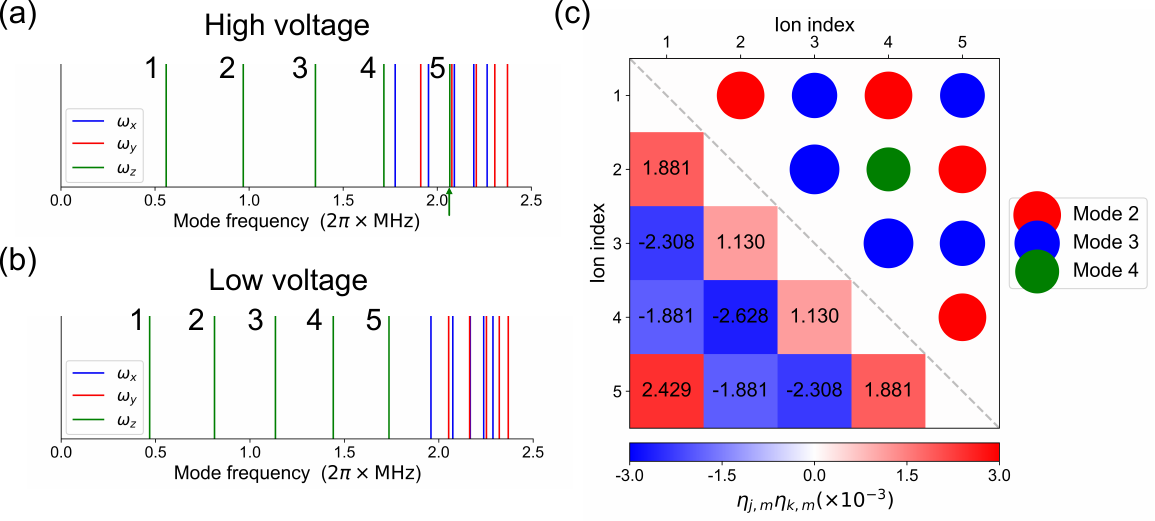}}
	\caption{
		(a) Frequency spectrum of axial motional modes under the high-voltage configuration, where the axial mode frequencies $\omega_z$ are $2\pi \times (0.550, 0.971, 1.356, 1.717, 2.065)\,{\rm MHz}$. 
		(b) Frequency spectrum of axial motional modes under the low-voltage configuration, yielding mode frequencies $\omega_z$  of $2\pi \times (0.455, 0.814, 1.140, 1.444, 1.733)\,{\rm MHz}$. The differences in the voltage settings alter both the trap potential and the inter-ion spacing, thereby modifying the mode structure.
		(c) Summary of axial mode selection and coupling strengths for all ion pairs. Each circle represents a two-qubit gate: the color denotes the axial mode used, the size reflects the experimental Rabi rate of the bichromatic laser field, and the value to the left indicates the product of Lamb-Dicke parameters $\eta_{j,m}\eta_{k,m}$, which determines the intrinsic coupling strength. Mode selection is guided by both coupling optimization and physical properties of the modes, such as suppression of Kerr-type nonlinearities in the low-voltage setting.}
	
	\label{figs1}
\end{figure*}

In this experiment, we employ two axial configurations with different voltage settings. Under the high-voltage setting, the axial mode frequencies are $2\pi \times (0.550, 0.971, 1.356, 1.717, 2.065)\,{\rm MHz}$, as shown in Fig.\,\ref{figs1}(a), and the  relative positions of the five ions at equilibrium are $(-11.425, -5.404, 0, 5.404, 11.425)\,\upmu{\rm m}$. Under the low-voltage setting, the axial mode frequencies are $2\pi \times (0.455, 0.814, 1.140, 1.444, 1.733)\,{\rm MHz}$, as shown in Fig.\,\ref{figs1}(b), and the equilibrium positions become $(-12.839, -6.089, 0, 6.089, 12.839)\,\upmu{\rm m}$. The gates between different ion pairs in the experiment are implemented by applying laser fields near resonant with selected axial modes, with optimal coupling strengths of choice to minimize gate duration. In Fig.\,\ref{figs1}(c), the color of each circle represents the axial mode used for a specific ion pair,  the circle size reflects the experimental Rabi rate of the dichromatic laser field, the value displayed on the bottom left part corresponds to the coupling strength. Another factor considered in mode selection is the Kerr-type nonlinear coupling between the axial and radial modes, inducing frequency shifts on the axial mode frequencies depending on the temperature of the radial modes. For our demonstration, the gates utilizing modes 3 and 4 are applied with the low-voltage setting, while the gates involving mode 2 are applied with the high-voltage setting. During the verification process, we find comparable gate fidelity can also be achieved under the low-voltage setting, as confirmed via numerical simulations .

%As discussed in Section~I, the geometric phase accumulated after two loops of the gate operation is $\theta$, which indicates that the product $\eta_{j,m}\eta_{k,m}$ determines the intrinsic coupling strength between ion pairs. This serves as a primary criterion for selecting different axial modes for different ion pairs in the experiment. In Fig.\,\ref{figs1}(c), the color of each circle represents the axial mode used for a specific ion pair, and the circle size reflects the experimental Rabi rate of the dichromatic laser field. and the value displayed on the left corresponds to the coupling strength. Another factor considered in mode selection is the physical characteristics of each axial mode. For gates utilizing modes 3 and 4, we apply the low-voltage setting, which reduces the influence of Kerr-type nonlinear coupling in radial modes and yields stronger effective coupling compared to the high-voltage configuration. For gates involving mode 2, we use the high-voltage setting; however, numerical simulations confirm that comparable gate fidelity can also be achieved under the low-voltage setting.

\section{III. Details of Individual Qubit Control}

\begin{figure*}
	\centering
	\makebox[\textwidth][c]{\includegraphics[width=16cm]{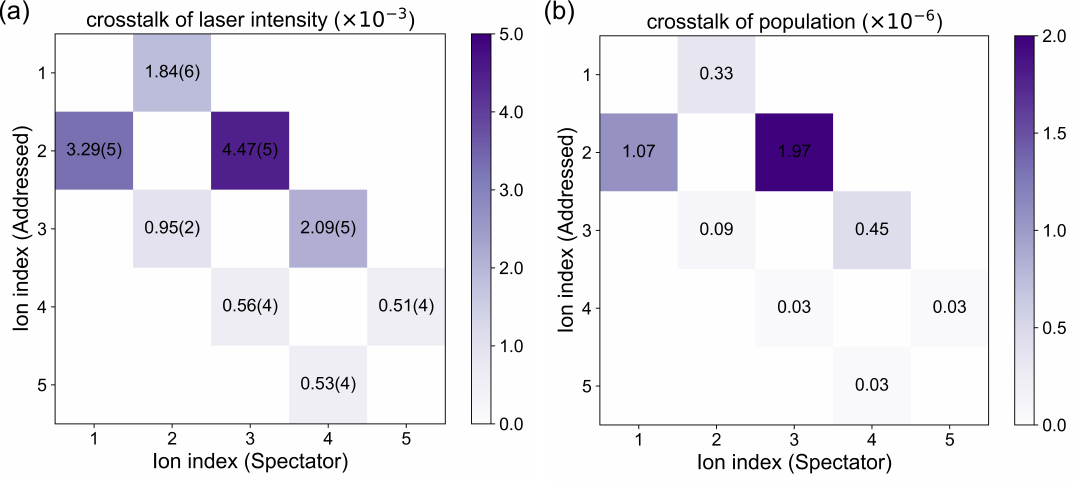}}
	\caption{
		(a) Measured intensity crosstalk of the 397~nm individual addressing beams. The crosstalk is inferred from ac Stark shifts on the $\ket{0} \leftrightarrow \ket{1}$ transition using a Ramsey interference method, where one of the neighboring ions is illuminated during the Ramsey delay. The observed asymmetry is attributed to optical aberrations in the beam delivery system. 
		(b)    Population crosstalk on the 397~nm Raman transition $\ket{g} \leftrightarrow \ket{1}$. This is derived by combining the ac Stark shift data from the $\ket{g} \leftrightarrow \ket{1}$ transitions and intensity crosstalk in (a), to estimate the maximum off-resonant population excitation induced by a 397~nm Raman $\pi$ pulse during state preparation. The low crosstalk observed is consistent with the modest ac Stark shift ($\sim 35\,\mathrm{kHz}$) relative to the Rabi rate, though it still contributes to Raman operation errors due to intensity fluctuations. 
	}
	\label{figs2}
\end{figure*}

To evaluate the performance of individual qubit control, we measure the addressing crosstalk of the laser intensity distribution under the high-voltage setting. Since the AC Stark shift is proportional to laser intensity~\cite{foot}, we determine the intensity crosstalk experienced by an ion in the chain by measuring the AC Stark shift using a Ramsey method on the $\ket{0} \leftrightarrow \ket{1}$ transition, where one of the adjacent ions is addressed during the Ramsey delay. The results are shown in Fig.\,\ref{figs2}(a), where the observed asymmetry is likely attributable to optical aberrations. 
Combined with the AC Stark shift data measured using the Ramsey method on the $\ket{g} \leftrightarrow \ket{1}$ transition, we calculate the maximum amplitude of off-resonant oscillations, as shown in Fig.\,\ref{figs2}(b). This represents the population crosstalk induced by the addressed 397~nm $\pi$ pulse used in the state preparation process, as discussed in the main text. We note that the typical AC Stark shift on the $\ket{g} \leftrightarrow \ket{1}$ transition under experimental conditions is approximately $35\,{\rm kHz}$, which is comparable to the Rabi rate of the same transition. This explains the small observed population crosstalk, while also contributing to 397~nm Raman operation errors during initial state preparation due to intensity fluctuations in the addressing beam. Active laser intensity stabilization, which would suppress this source of error, is scheduled in future upgrade plans.

\section{IV. Numerical Simulation for Error Estimation}

To model the error of the MS gate, we characterize the noise and decoherence properties of the system. The spin dephasing time is measured as $T_2^{*} = 4.3~{\rm ms}$ under Gaussian noise using the standard Ramsey method on the $\ket{0} \leftrightarrow \ket{1}$ transition. Additionally, Carr–Purcell–Meiboom–Gill (CPMG)~\cite{CPMG} experiments with sequence lengths $L = 9$ and $L = 21$ confirm the absence of prominent noise peaks in the frequency range from $5~{\rm kHz}$ to $50~{\rm kHz}$. These results indicate that spin dephasing is predominantly affected by low-frequency noise, which we model using shot-to-shot sampling.

The decoherence parameters for the axial motional modes (hereafter also referred to as phonons) used in the experiment are summarized in Table~\ref{tab1}. Mode heating is characterized by fitting the blue sideband evolution and extracting the mean phonon number at various waiting times after sideband cooling (SBC). Phonon Ramsey experiments with and without a phonon echo pulse are performed using the method in Ref.~\cite{phne}, and the data is fitted assuming Gaussian-form noise. The suppression of dephasing by the echo pulse suggests that phonon decoherence also originates primarily from low-frequency noise, which is likewise modeled via shot-to-shot sampling.

Rabi rate fluctuations are modeled by shot-to-shot sampling with a standard deviation of $0.8\%$, arising from two contributions: direct laser intensity fluctuations of the 729~nm beam ($0.25\%$) and the Debye–Waller effect due to nonzero initial phonon occupation ($0.55\%$). The error associated with the decay of the metastable $D_{5/2}$ state is obtained by integrating the gate evolution over time under a fixed decay rate of $\tau = 1/1168~{\rm ms}^{-1}$. The error due to off-resonant excitation is calculated by considering both the carrier transition and the two adjacent axial modes using experimental parameters.

In addition to the above error sources—also listed in the main text—we estimate the contributions of other possible imperfections, such as the AC Stark shift, imbalance between the two frequency components of the 729~nm bichromatic field, and Rabi rate mismatch between the two ions. These effects are found to be negligible under our experimental conditions.

\begin{table}[h]
	\centering
	\caption{Decoherence parameters of the axial motional modes.} \begin{tabular}{p{6cm}|p{3cm}<{\centering}p{3cm}<{\centering}p{3cm}<{\centering}}
		\hline
		& Mode 2 & Mode 3 & Mode 4 \\
		\hline
		Heating rate (quanta/s) & 29(23) & 104(16) & 5(18) \\
		Phonon dephasing time (ms) & 1.5 & 0.9 & 0.5 \\
		Phonon dephasing time (echo) (ms) & 13.0 & 8.8 & 11.2 \\
		\hline
	\end{tabular}
	\label{tab1}
\end{table}

\section{V. Post-Selection Method and Readout Error Analysis}

We apply post-selection to alleviate the effect of imperfect state preparation, for more accurate evaluation of the gate fidelity \cite{Gae, Qud1, Qud2,CSh,jeff}, mentioned in the main text. In the following, we first describe such a process in detail, and then analyze various imperfections in the post-selection process.

%, as errors that drive population out of the computational space are comparable to or even exceed the intrinsic MS gate error. 
Because the detection scheme is based on collecting fluorescence from the 397~nm cycling transition between $S_{1/2}$ and $P_{1/2}$, spin states $\ket{1}$ and $\ket{g}$ both appear ``bright,'' while all $D_{5/2}$ sublevels---including $\ket{0}$ and auxiliary levels $\ket{a_1}$ and $\ket{a_2}$---appear ``dark''. Therefore, even in the ideal case of zero technical error during state detection, a single detection event cannot definitively determine whether the ion remains within the computational space. To resolve this, our post-selection strategy uses two consecutive detection events in a single shot to identify spin states more reliably~\cite{Gae, Qud1, Qud2}.
The correspondence between spin states and two fluorescence detection results is summarized in Table~\ref{tab2}. Notably, both $\ket{1}$ and $\ket{g}$ give ``bright--dark'', and $\ket{0}$ yields ``dark--bright''. The auxiliary states $\ket{a_1}$ and $\ket{a_2}$ appear as ``dark--dark'', allowing us to effectively group them as a single state $\ket{a}$. Thus, the measurement outcomes classify into three categories: $\ket{0}$ (dark--bright), $\ket{1}$ or $\ket{g}$ (bright--dark), and $\ket{a}$ (dark--dark). This classification enables us to design a reliable post-selection protocol that screens out corrupted shots.

Consider ion pair (1,\,2) as an example. After spin initialization and addressed 397~nm Raman $\pi$ pulses, the five-ion state becomes
\[
(n_1\ket{1}_1 + m_1\ket{g}_1)(n_2\ket{1}_2 + m_2\ket{g}_2) \otimes \ket{ggg}_{345},
\]
where $\left|n_i\right|^2 + \left|m_i\right|^2 = 1$ and $\left|n_i/m_i\right|^2 > 15$ thanks to high selectivity and negligible population crosstalk ($< 2.0 \times 10^{-6}$). Two sequential shelving pulses transfer $\ket{g}$ to  $\ket{a_2}$ and $\ket{a_1}$, leaving most population in $\ket{a}$ and a small residual in $\ket{g}$ ($< 2.5 \times 10^{-5}$ per ion). Off-resonant excitation of $\ket{1}$ to $D_{5/2}$ is also negligible ($< 10^{-5}$), so the spin state becomes
\[
(n_1\ket{1}_1 + m_1\ket{a}_1)(n_2\ket{1}_2 + m_2\ket{a}_2) \otimes \ket{aaa}_{345},
\]
which expands to:
\[
(n_1 n_2 \ket{11}_{12} + n_1 m_2 \ket{1a}_{12} + m_1 n_2 \ket{a1}_{12} + m_1 m_2 \ket{aa}_{12}) \otimes \ket{aaa}_{345}.
\]
Only the $\ket{11}$ component participates in the MS gate, while all others are either non-interacting or invalid. Crucially, these unwanted components can be identified by their detection signatures. 
%For instance, $\ket{a}$ gives two ``dark'' outcomes and can be filtered out reliably.
We thus define the post-selection criteria using the two detection results: (i) for the two target ions, we retain a shot only if one detection shows ``bright'' and the other ``dark'', indicating presence in the computational basis ($\ket{0}$ or $\ket{1}$); (ii) for the spectator ions, we retain a shot only if both detections return ``dark'', confirming their shelving into $\ket{a}$. This ensures that only computational-state components evolve under the MS gate and all interactions involving leaked states are excluded.
%Assuming that preparation errors are independent of gate errors, this method isolates the intrinsic two-qubit gate fidelity. The normalized populations $P_{00}$, $P_{01}$, $P_{10}$, and $P_{11}$ derived from post-selected shots thus provide an accurate estimate of the gate performance, free from the influence of unrelated errors.

\begin{table}[h]
	\centering
	\caption{Correspondence between spin states and two detection results} 
	\begin{tabular}{p{5cm}|p{4cm}<{\centering}p{4cm}<{\centering}}
		\hline
		Spin states & First detection & Second detection\\
		\hline
		$\ket{0}$ & Dark & Bright\\
		$\ket{1}$ & Bright & Dark\\
		$\ket{g}$  & Bright & Dark\\
		$\ket{a_1}$ & Dark & Dark\\
		$\ket{a_2}$ & Dark & Dark\\
		\hline
	\end{tabular}
	\label{tab2}
\end{table}

%The readout process in the experiment also introduces errors. By analyzing these errors, we can further suppress them to better reflect the true gate fidelity. 
Readout errors (see Fig. 2(b) in main text) will distort the measured populations after post-selection,  $P_{xy}^{\mathrm{exp}}$, compared to the ideal probabilities $P_{xy}$ for the two target ions (where $x,y \in \{0,1\}$). Following Ref.~\cite{CSh}, We characterize how errors at each step of the readout sequence affect the extracted fidelity.
%Before post-selection, each experiment yields two detection results per ion, and for a 5-ion chain, this results in $3^5 = 243$ possible detection outcome patterns. However, after post-selection, only a small subset of those outcomes—corresponding to valid configurations of target and spectator ions—are retained. 
%Specifically, we project the 5-ion state onto the effective measurement basis
% After the MS gate, the system is in the subspace
% $\{\ket{00}, \ket{01}, \ket{10}, \ket{11}\}_{12} \otimes \ket{aaa}_{345}$, which yields four relevant outcomes for the two-qubit state. 
Note that if the readout procedure were ideal, then by repeating the experiment $N$ times and retaining $N_{\text{tot}}$ post-selected shots, we would extract the probabilities as $P_{xy} = N_{xy}/N_{\text{tot}}$ with statistical uncertainty $\Delta P_{xy} = \sqrt{P_{xy}(1 - P_{xy}) / N_{\text{tot}}}$ from binomial statistics.
%In reality readout errors distort both $N_{xy}$ and $N_{\text{tot}}$, leading to experimentally observed probabilities $P_{xy}^{\mathrm{exp}} \neq P_{xy}$. 
The influence of readout errors on post-selection could be classified into three types:
First, errors that change outcomes within the computational subspace $\{\ket{00}, \ket{01}, \ket{10}, \ket{11}\}_{12} \otimes \ket{aaa}_{345}$, affecting $N_{xy}$ but not $N_{\text{tot}}$.
Second, errors that induce leakage from valid computational-basis states, reducing both $N_{xy}$ and $N_{\text{tot}}$.
Third, errors that cause invalid (non-computational) states to be mistakenly retained, increasing both $N_{xy}$ and $N_{\text{tot}}$.
The third situation involves rare coincidences (e.g., leakage into $\ket{a}$ during state preparation, combined with readout errors returning a false detection signature). Since the leakage probability is $\lesssim 11\%$ and the readout error causing false detection signature is $\lesssim 1\%$, the combined effect is $<10^{-3}$ and can be safely neglected.
We now describe how each step in the readout sequence maps ideal probabilities $P_{xy}$ to experimentally observed ones $P_{xy}^{\mathrm{exp}}$, through distortion operations denoted $\mathcal{E}_i$. Each $\mathcal{E}_i$ models a specific readout step assuming all other steps are ideal. 

\begin{table}[htb]
	\centering
	\caption{Discard rates due to post selection for all ten ion pairs.} 
	\begin{tabular}{|p{2cm}<{\centering}||p{3cm}<{\centering}p{3cm}<{\centering}|p{3cm}<{\centering}p{3cm}<{\centering}|}
		\hline
		ion pair & population & discard rate & parity contrast & average discard rate \\
		\hline
		(1, 2) & 0.998(2) & 0.067 & 0.987(4) & 0.076\\
		(1, 3) & 0.998(2) & 0.065 & 0.989(3) & 0.065\\
		(1, 4) & 0.989(3) & 0.043 & 0.983(4) & 0.104\\
		(1, 5) & 0.997(2) & 0.077 & 0.969(5) & 0.094\\
		(2, 3) & 0.999(1) & 0.067 & 0.977(5) & 0.091\\
		(2, 4) & 0.992(3) & 0.088 & 0.966(5) & 0.084\\
		(2, 5) & 0.986(4) & 0.075 & 0.994(3) & 0.073\\
		(3, 4) & 0.993(3) & 0.105 & 0.980(4) & 0.089\\
		(3, 5) & 0.992(3) & 0.093 & 0.978(4) & 0.063\\
		(4, 5) & 0.997(2) & 0.059 & 0.987(4) & 0.065\\
		\hline
	\end{tabular}
	\label{tab3}
\end{table}

\noindent\textbf{Step 1: $\pi/2$ pulse with varying phase (parity analysis only).}  
This step is only applied during parity measurements. Since the infidelity of the $\pi/2$ pulse is estimated to be much smaller than the two-qubit gate error, its contribution is negligible. Therefore, we take $\mathcal{E}_1$ as the identity operation in our error model.

\noindent\textbf{Step 2: First fluorescence detection.}
This step can cause false ``bright'' detection of $\ket{0}$ due to decay from $D_{5/2}$, quantified by $\varepsilon_D \approx 0.2\%$ for a single ion. This error distorts the population count. The mapping $\mathcal{E}_{2}$ is described by:
\[
\begin{bmatrix}
	P^{\textrm{exp}}_{00}\\
	P^{\textrm{exp}}_{01}\\
	P^{\textrm{exp}}_{10}\\
	P^{\textrm{exp}}_{11}
\end{bmatrix}=\begin{bmatrix}
	(1 - \varepsilon_D)^2 & 0 & 0 & 0 \\
	(1 - \varepsilon_D)\varepsilon_D & 1 - \varepsilon_D & 0 & 0 \\
	\varepsilon_D(1 - \varepsilon_D) & 0 & 1 - \varepsilon_D & 0 \\
	\varepsilon_D^2 & \varepsilon_D & \varepsilon_D & 1
\end{bmatrix}
\begin{bmatrix}
	P_{00}\\
	P_{01}\\
	P_{10}\\
	P_{11}
\end{bmatrix}.
\]

\noindent\textbf{Step 3: Doppler and EIT cooling.}
During cooling ($\sim$ 6.5 ms total), $\ket{0}$ can decay into $S_{1/2}$ and be pumped into $\ket{1}$ before the next detection, thereby violating post-selection and reducing $N_{\text{tot}}$. This leakage is modeled by:
\[
\mathcal{E}_{3}:\left\{
\begin{aligned}
	P_{00}^{\mathrm{exp}} &= P_{00}(1 - \varepsilon_C)^2 / A_3 \\
	P_{01}^{\mathrm{exp}} &= P_{01}(1 - \varepsilon_C) / A_3 \\
	P_{10}^{\mathrm{exp}} &= P_{10}(1 - \varepsilon_C) / A_3 \\
	P_{11}^{\mathrm{exp}} &= P_{11} / A_3,
\end{aligned}
\right.
\]
with $A_3 = P_{00}(1 - \varepsilon_C)^2 + P_{01}(1 - \varepsilon_C) + P_{10}(1 - \varepsilon_C) + P_{11}$. Here, $\varepsilon_C \approx 0.55\%$ is the estimated decay probability during cooling.

\noindent\textbf{Step 4: Optical pumping to $\ket{1}$.}
Failure to pump from $\ket{g}$ to $\ket{1}$ leads to incorrect shelving and false ``bright'' detection. The corresponding mapping is:
\[
\mathcal{E}_{4}:\left\{
\begin{aligned}
	P_{00}^{\mathrm{exp}} &= P_{00} / A_4 \\
	P_{01}^{\mathrm{exp}} &= P_{01}(1 - \varepsilon_P) / A_4 \\
	P_{10}^{\mathrm{exp}} &= P_{10}(1 - \varepsilon_P) / A_4 \\
	P_{11}^{\mathrm{exp}} &= P_{11}(1 - \varepsilon_P)^2 / A_4,
\end{aligned}
\right.
\]
where $\varepsilon_P < 0.1\%$ and $A_4 = P_{00} + P_{01}(1 - \varepsilon_P) + P_{10}(1 - \varepsilon_P) + P_{11}(1 - \varepsilon_P)^2$.

\noindent\textbf{Step 5: Global 729~nm $\pi$ pulse.}
Errors in this $\pi$ pulse (denoted $\varepsilon_\pi$) yield same-outcome patterns (e.g., bright-bright) for each target ion and violate post-selection. However, the form of the resulting mapping,
\[
\mathcal{E}_5:
P_{xy}^{\mathrm{exp}} = P_{xy}(1 - \varepsilon_\pi)^2 / A_5,
\]
shares a common factor $A_5=(1 - \varepsilon_\pi)^2$, meaning that the distortion cancels in normalized populations.
%, so $\mathcal{E}_5$ reduces to an identity operation.

\noindent\textbf{Step 6: Second fluorescence detection.} This step has the same detection error $\varepsilon_D$ as in step 2,
\[
\mathcal{E}_6:\left\{
\begin{aligned}
	P_{00}^{\mathrm{exp}} &= P_{00} / A_6 \\
	P_{01}^{\mathrm{exp}} &= P_{01}(1 - \varepsilon_D) / A_6 \\
	P_{10}^{\mathrm{exp}} &= P_{10}(1 - \varepsilon_D) / A_6 \\
	P_{11}^{\mathrm{exp}} &= P_{11}(1 - \varepsilon_D)^2 / A_6,
\end{aligned}
\right.
\]
with $A_6 = P_{00} + P_{01}(1 - \varepsilon_D) + P_{10}(1 - \varepsilon_D) + P_{11}(1 - \varepsilon_D)^2$.

%By composing all mappings $\mathcal{E}_i$, we propagate the ideal two-qubit state populations to the experimentally observed values, $P_{xy} \rightarrow P_{xy}^{\mathrm{exp}}$. 
As the probability of multiple ($>1$) errors occurring is small (on the order of $10^{-5}$), 
only cases where an error occurs once in a single step are considered. 
%Under the reasonable approximation that the occurrence of multiple coinciding errors is negligible , 
%The reduction in $N_{\text{tot}}$ can be attributed to specific steps individually. This sequential error model thus provides an accurate and reliable estimate within experimental uncertainty.
%Since the mapping from $P_{xy}$ to $P_{xy}^{\mathrm{exp}}$ depends on $P_{xy}$ themselves, 
%we analyze the distortion specifically 
Consider ideal Bell states 
%in a high-fidelity gate, 
including $\left(\ket{00} + e^{i\varphi} \ket{11} \right)/\sqrt{2}$ and $\left(\ket{01} + e^{i\varphi} \ket{10} \right)/\sqrt{2}$. %where the phase $\varphi$ is irrelevant for our analysis. 
The former describes the Bell state directly after entangling gate operations (or the maximum-parity state after the analysis $\pi/2$ pulse), 
%and under the first-order approximation 
neglecting terms involving products of error probabilities, we find:
\[
P_{00}^{\mathrm{exp}} = P_{00} + \frac{\varepsilon_P - \varepsilon_C - \varepsilon_D}{2}, \quad
P_{11}^{\mathrm{exp}} = P_{11} + \frac{\varepsilon_C - \varepsilon_P - \varepsilon_D}{2}.
\]
The latter Bell state corresponds to the minimum-parity state after the analysis pulse, for which the error propagation gives:
\[
P_{00}^{\mathrm{exp}} = P_{00}, \quad P_{11}^{\mathrm{exp}} = P_{11} + \varepsilon_D.
\]
Using the fidelity formula $
\mathcal{F}_{\text{Bell}} = \frac{1}{2} \left( P_{00} + P_{11} + C \right) 
$, the observed population sum $P_{00}^{\mathrm{exp}} + P_{11}^{\mathrm{exp}} = P_{00} + P_{11} - \varepsilon_D$, and the measured parity contrast becomes $C^{\mathrm{exp}} = C - 2\varepsilon_D$. Consequently, the  Bell state fidelity is underestimated by $3\varepsilon_D / 2$ due to readout errors. The intrinsic MS gate error should thus be corrected by subtracting this factor. 
%Notably, other types of errors primarily deform the underlying populations $P_{xy}$ without significantly altering the calculated fidelity.
Detailed numerical results, including discard rates for all ten ion pairs, are summarized in Table~\ref{tab3}. The discard rates are consistent with the two-ion preparation infidelity, where single ion state preparation infidelity is estimated to be approximately on the 5\% level.

\end{document}